\documentclass{sig-alternate}
\usepackage[utf8]{inputenc}
\usepackage{enumitem}

\begin{document}






%

\title{Summary of the 3\raisebox{0.5em}{\large{\textsf{\textbf{rd}}}} 
International Workshop on Requirements Engineering and Testing (RET 2016)}
\subtitle{[Co-located with REFSQ 2016]}
%
%
%
%
%

\numberofauthors{6} 
%
\author{
%
%
\alignauthor
Michael Unterkalmsteiner\\
       \affaddr{Software Engineering Research Lab Sweden}\\
       \affaddr{Blekinge Institute of Technology}\\
       \affaddr{Karlskrona, Sweden}\\
       \email{mun@bth.se}
\alignauthor
Gregory Gay\\
       \affaddr{University of South Carolina}\\
       \affaddr{Columbia, SC, USA}\\
       \email{greg@greggay.com}
\alignauthor 
Michael Felderer\\
       \affaddr{University of Innsbruck}\\
       \affaddr{Innsbruck, Austria}\\
       \email{michael.felderer@uibk.ac.at}
\and  
\alignauthor 
Elizabeth Bjarnason\\
       \affaddr{Lund University}\\
       \affaddr{Lund, Sweden}\\
       \email{elizabeth.bjarnason@cs.lth.se}
\alignauthor 
Markus Borg\\
       \affaddr{SICS Swedish ICT AB}\\
       \affaddr{Lund, Sweden}\\
       \email{markus.borg@sics.se}
\alignauthor 
Mirko Morandini\\
       \affaddr{Fondazione Bruno Kessler}\\
       \affaddr{Trento, Italy}\\
       \email{morandini@fbk.eu}
}

\maketitle
\begin{abstract}
The RET (Requirements Engineering and Testing) workshop series provides a 
meeting point for researchers and practitioners from the two separate fields of 
Requirements Engineering (RE) and Testing. The goal is to improve the 
connection and alignment of these two areas through an exchange of ideas, 
challenges, practices, experiences and results. The long term aim is to build a 
community and a body of knowledge within the intersection of RE and Testing, 
i.e. RET. The 3rd workshop was held in co-location with REFSQ 2016 in 
Gothenburg, Sweden. The workshop continued in the same interactive vein as the 
predecessors and included a keynote, paper presentations with ample time for 
discussions, and panels. In order to create an RET knowledge base, this 
cross-cutting area elicits contributions from both RE and Testing, and from both
researchers and practitioners. A range of papers were presented from
short positions papers to full research papers that cover connections
between the two fields. 
\end{abstract}

\category{D.2.1}{Requirements\,/\,Specifications}{}
\category{D.2.4}{Software / Program Verification}{}
\category{D.2.5}{Testing and Debugging}{}

\terms{Management, Documentation, Human Factors, Verification}

\keywords{requirements engineering, testing, coordination, alignment}

\section{Introduction}\label{sec:intro}
The main objective of the RET workshop series is to explore, characterize and 
understand the interaction of Requirements Engineering (RE) and Testing, both 
in research and industry, and the challenges that result from this interaction. 
The workshop provides a forum for exchanging experiences, ideas and best 
practices to coordinate RE and testing. A primary goal of this exchange is to 
enable and provide incentives for research that crosses research areas and is 
relevant for industry. Towards this end, RET invites submissions exploring how 
to coordinate RE and Testing, including practices, artifacts, methods, 
techniques and tools. Submissions on softer aspects like the communication 
between roles in engineering processes are also welcome.

RET 2016 accepted technical papers with a maximum length of 15 pages presenting 
research results or industrial practices and experiences related to the 
coordination of RET, as well as position papers with a maximum length of 6 
pages introducing challenges, visions, positions or preliminary results within 
the scope of the workshop. Experience reports and papers on open challenges in 
industry were especially welcome.

RET 2016 accepted four technical papers and one position paper. The workshop 
was visited by 17 participants and the proceedings are available 
online~\cite{bjarnason_joint_2016}.

\section{Organization}
The 3rd International Workshop on Requirements Engineering and Testing (RET 
2016) was held on March 14, 2016, and was co-located with the 22nd 
International Working Conference on Requirements Engineering: Foundation for 
Software Quality (REFSQ 2016). The website for the workshop is available 
online\footnote{http://ret.cs.lth.se/16}. The workshop was organized by Michael 
Unterkalmsteiner (Blekinge Institute of Technology) as general chair, Gregory 
Gay (University of South Carolina) and Michael Felderer (University of 
Innsbruck) as program co-chairs, as well as, Elizabeth Bjarnason (Lund 
University), Markus Borg (SICS Swedish ICT AB) and Mirko Morandini (Fondazione 
Bruno Kessler) as co-chairs.

\section{Program Summary}
The program of RET 2016 comprised of an introductory part with a keynote, a 
discussion on the past and future of RET, and two paper presentation sessions 
followed by panels with the paper presenters.

After a welcome note, Baldvin Gislason Bern (R\&D Expert at Axis Communications 
AB) gave an invited talk entitled ``Tests as requirements - Why we don't do 
requirements at Axis''. He illustrated why and how Axis, a company with 
approximately 100 different products, can be successful with a team of 120 
test engineers and very little resources dedicated at requirements 
engineering and management. A major factor is the technology-driven domain the 
company is operating in, allowing product managers to push new technologies to 
the market, collecting data from customers and then acting upon this feedback 
to improve future iterations of the product. They key take-way idea from the 
presentation is that knowledge within a company on a technology/product/market 
changes over time, requiring flexible strategies for product development. With 
a novel technology, little knowledge and experience exist and requirements are 
unclear. The role of testing is to explore limits to gain knowledge. Therefore, 
test cases document decisions, as opposed to requirements which would document 
intentions. Knowledge that stems from testing the product is documented in test 
cases, rendering them a living documentation that is actively used and 
maintained over time. As the knowledge on technology in Axis matures and to be 
able to maintain a competitive edge, the company is adapting its strategy 
towards value driven product development. This will put customer feedback 
in the center, and, as a new source of requirements, drive product development.

\begin{figure}
	\centering
	\includegraphics[width=1\linewidth]{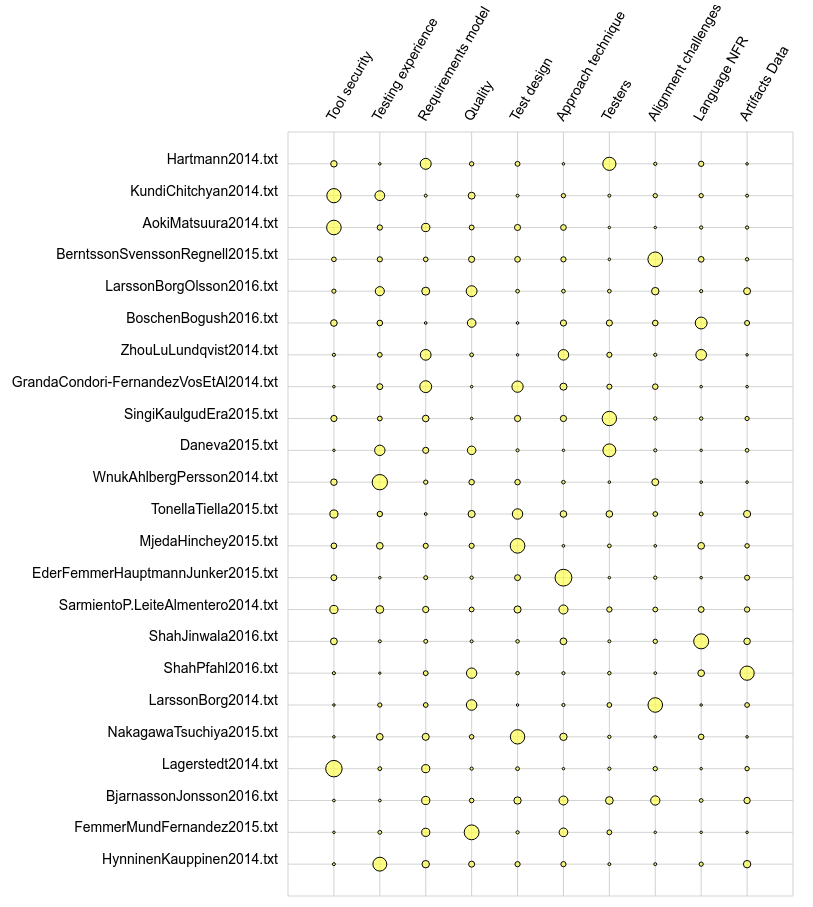}
	\caption{Topic model of the papers accepted at RET 2014-2016}
	\label{fig:topic_model}
\end{figure}

The keynote was followed by a panel on future industry needs with respect to 
coordinating requirements engineering and testing. To kick-off the discussion, 
the workshop chair presented a thematic summary of the three instances of 
workshop in 2014, 2015 and 2016, generating a topic model from the 23 abstract 
that were accepted and presented in total at the workshop. He illustrated the 
predominant themes with Serendip~\cite{alexander_serendip:_2014}, a 
visualization tool for topic models (see Figure~\ref{fig:topic_model}). The 
rows represent the papers presented at RET in three years. The columns 
represent the identified topics\footnote{The number of topics, 10, is a 
required parameter when generating the model and was set rather arbitrarily. 
However, 10 topics seemed to be enough to provide some differentiation between 
papers and not too much to be too fine-grained. Most of the topics were rather 
easy to label, based on the most frequent terms per topic.}. The size of the 
circle on the crossing between article and topic represents the probability 
that the document was generated by the terms that represent the respective 
topic. The predominant topics at the respective workshop instances were:
\begin{itemize}
	\item RET 2014: Tools, Security Requirements, System Testing, Experience, 
	Development and Models
	\item RET 2015: Test Design, Testers, Quality
	\item RET 2016: Language, Quality, Artifacts and Data
\end{itemize}
This result suggests that the three workshops were driven by different themes, 
quality being a commonality between 2015 and 2016.
In Figure~\ref{fig:topic_model}, the topics are ordered from left to right, by 
the total proportion. The ``tool'' topic predominates, followed by ``testing 
experience'' and ``requirements model''. This result suggests that the accepted 
papers are thematically in line with the goals of RET (see 
Section~\ref{sec:intro}), at least from the perspective of the used vocabulary 
in the abstracts.
Based on these results and the preceding keynote, the panelists consisting of 
two researchers and three industry participants discussed future research 
avenues for RET. A common interest seemed to be the effective use of customer 
feedback to drive product development. This approach is quite feasible as one 
panelist from the mobile app domain observed, may however be impractical in 
other domains such as the automobile industry where feedback cycles with 
customers are more difficult to establish.

The panel was followed by the first paper session themed quality requirements. 
The talk ``Testing Quality Requirements of a System-of-Systems in the Public 
Sector - Challenges and Potential Remedies'' identified five main challenges 
when testing quality requirements: (1) evolving RE documents while testing is 
planned and ongoing, (2) test managers need to understand the business side of 
the company, (3) quality requirements are not quantified or (4) prioritized, 
(5) difficulty to generate test data that exercises all operational states. 
These challenges were matched with solution proposals from the scientific 
literature.

The talk ``Evaluating and Improving Software Quality Using Text Analysis 
Techniques - A Mapping Study'' identified 81 primary studies. The most frequent 
application of text analysis techniques was in defect management (bug 
classification and severity assignment), followed by requirements engineering 
(concept extraction). The most common data sources are bug reports and 
requirement documents. Interesting avenues for future research are to study 
mobile app reviews to understand software quality and to combine multiple data 
sources.

The talk ``Specification of Non-Functional Requirements: A Hybrid Approach'' 
proposed an approach to identify NFRs in natural language requirements using 
text processing techniques and ontologies, which are then modeled as use cases.
The approach has been illustrated in a case study with a proof-of-concept 
example.

The second paper session comprised of a position paper and a full technical 
paper. The talk ``Improving Project Coordination through Data Mining and 
Proximity Tracking'' proposed to analyze what project members work on and to 
support direct interaction based on common work related themes when individuals 
meet. Proximity tracking would also allow to identify communication patterns 
that could provide insights who collaborates with whom and when.

The talk ``Bridging the Gap between Natural Language Requirements and Formal 
Specifications'' proposed to use requirements boilerplates to support the 
formalization process of natural language requirements. The industry case study 
conducted at Airbus illustrated how a requirements quality tool was used to 
extract semantics from boilerplates to semi-automatically generate formal 
requirements specifications. 

\section{Future}
We plan to organize the workshop again next year since the topic attracted 
interested from both industry and academia. Our aim is to organize RET 2017 
co-located with the 25th International Requirements Engineering Conference (RE 
2017) in Portugal. If the workshop is accepted, the expected date for paper 
submissions is in June 2017.

\section{Acknowledgments}
We want to thank the participants of the workshop and all the authors of
submitted papers for their important contribution to the event. In
addition, we want to thank the organizers of the 22nd International Working 
Conference on Requirements Engineering: Foundation for Software Quality (REFSQ 
2016) and the members of the program committee:
\begin{itemize}[leftmargin=*]
	\itemsep-0.3em
	\item [] \textbf{Armin Beer}, Beer Test Consulting, Austria
	\item [] \textbf{Ruzanna Chitchyan}, Leicester University, UK
	\item [] \textbf{Nelly Condori-Fernandez}, PROS Research Centre, Spain
	\item [] \textbf{Robert Feldt}, Blekinge Institute of Technology, Sweden
	\item [] \textbf{Henning Femmer}, Technische Universität München, Germany
	\item [] \textbf{Vahid Garousi}, Hacettepe University, Turkey
	\item [] \textbf{Joel Greenyer}, University of Hannover, Germany
	\item [] \textbf{Andrea Herrmann}, Herrmann \& Ehrlich, Germany
	\item [] \textbf{Mike Hinchey}, Lero - the Irish Software Engineering 
	Research Centre, Ireland
	\item [] \textbf{Jacob Larsson}, Capgemini, Sweden
	\item [] \textbf{Annabella Loconsole}, Malmö University, Sweden
	\item [] \textbf{Alessandro Marchetto}, FIAT research, Italy
	\item [] \textbf{Cu Duy Nguyen}, University of Luxembourg, Luxembourg
	\item [] \textbf{Magnus C. Ohlsson}, System Verification, Sweden
	\item [] \textbf{Barbara Paech}, University of Heidelberg, Germany
	\item [] \textbf{Dietmar Pfahl}, University of Tartu, Estonia
	\item [] \textbf{Sanjai Rayadurgam}, University of Minnesota, USA
	\item [] \textbf{Giedre Sabaliauskaite}, Singapore University of Technology 
	and Design, Singapore
	\item [] \textbf{Hema Srikanth}, IBM, USA
	\item [] \textbf{Marc-Florian Wendland}, Fraunhofer FOKUS, Germany
	\item [] \textbf{Dongjiang You}, University of Minnesota, USA
	\item [] \textbf{Yuanyuan Zhang}, University College London, UK
\end{itemize}

%
\bibliographystyle{abbrv}
\bibliography{summary}  
%
\end{document}